\documentstyle[preprint,aps,epsf]{revtex}
\begin{document}
\draft

\title { Spectral properties of one dimensional 
insulators and superconductors }
\author{S. Sorella}
\address{
INFM and International School for Advanced Study\\
Via Beirut 4, 34013 Trieste, Italy}
\author{A. Parola}
\address{INFM and Istituto di Scienze Fisiche, 
Universit\'a di Milano, Via Lucini 3,
Como, Italy.}
\date{SISSA preprint: 19/96/CM/MB}
\maketitle
\begin{abstract}\widetext
Conformal field theory and Bethe ansatz are used to 
investigate the low energy features of the spectral function in
one dimensional models which exhibit a gap in the spin or in the charge 
excitation spectrum. Exotic behavior is found in the superconducting 
case, where the Green function displays momentum dependent Luttinger 
Liquid exponents. The predictions of the formalism are confirmed by 
Lanczos diagonalizations in the $tJ$ model up to 32 sites. These results 
may be relevant in connection to photoemission experiments in quasi one 
dimensional insulators or superconductors.
\end{abstract}
\widetext

In the last few years direct and inverse photoemission experiments have 
considerably improved, allowing for the accurate determination of momentum 
dependent energy spectra in low dimensional systems \cite{wells,aebi,quasi1d}.
A first principle interpretation of this class of experiments requires a
deep understanding of the effects of correlations on the electron (or hole) 
spectral function.
One dimensional (1D) metals \cite{haldane},
have been the subject of an intense theoretical effort which led to a
complete characterization of the long wavelength, low energy properties of
electron dynamics and to the actual calculation of the correlation 
exponents which appear in the electron Green function \cite{korepin,review}.
However, some doubt has been cast on the relevance of these calculations for
photoemission experiments because most of quasi 1D systems are close to 
density wave or superconducting instabilities which open a gap in the charge 
or spin spectrum \cite{quasi1d}. If the excitation spectrum is {\it fully 
gapped}, the Green function takes a free particle-like form in every
dimension. Instead, the effects of a {\it a single branch} of gapless 
excitations (either spin or charge) have not been addressed in detail before, 
probably because the prejudice prevailed that a system with a gap should 
display exponentially decaying correlation functions both in space and time.

In this Letter, we develop a microscopic theory for determining the low
energy properties of the spectral function in one dimensional correlated 
electron models with a gap either in the charge or in the spin channel. 
The main result of this work is that the presence of gapless excitations 
induces anomalous exponents in the  Green function, by 
a non trivial interaction with the extra electron (or hole) injected 
into the system. As a consequence, we generally find a spectral function
with singularities along lines in the $(k,\omega)$ plane. These 
singularities are characterized by critical exponents which possibly
depend on the {\it momentum} $k$ of the electron: Numerical diagonalizations
in the $tJ$ model at $J=2t$ fully confirm this picture providing quantitative
agreement with predictions based on conformal field theory and Bethe 
ansatz techniques. 

Spin and charge gaps are treated on the same footing 
in this Letter because they are believed to give rise to the same kind of
singularities in the spectral function, despite the quite distinct physical
nature of the state. In fact, it is possible to build specific models
where the two regimes are mapped one onto the other: A well known
example is the negative $U$ Hubbard model at zero magnetization which 
is the prototype of one dimensional ``superconductors" 
i.e. 1D models with spin gap and quasi off diagonal long range order. 
This model, via a particle-hole transformation, is mapped into the 
half filled, positive $U$ system with {\it non vanishing magnetization}
which, on the contrary, is a Mott insulator characterized by a charge gap.
The full Green function remains unaltered by particle-hole transformation 
leading to the same photoemission spectra in the two cases. In the following,
we will explicitly deal only with the repulsive, half filled case, 
at arbitrary magnetization. The results will however hold both for 
one dimensional insulators and ``superconductors", being related only 
to the presence of a branch of gapless excitations in the spectrum.

The quantity which we are going to investigate is the hole Green function:
\begin{equation} \label{holegreen}
G(p,t)=
i\,<\Psi_0|\,c^{\dagger}_{p,\sigma}\,e^{-i t (H-E_0-i\delta)}\,c_{p,\sigma}\,
|\Psi_0>\,\theta(t) 
\end{equation}
where $|\Psi_0>$ ($E_0$) is the ground state (energy) 
of the system with no holes
and  $\theta(t)$ is the step function.
Due to spin-charge decoupling, the total energy $E$ and momentum $p$
are naturally written as a sum of a holon term $\epsilon_h(k)$ and a
spinon contribution $\epsilon_s(Q)$ with $p=k+Q$.
At long wavelengths it is known that holons and spinons behave as independent
particles whose dynamics is governed by two commuting hamiltonians: $H_c$ and
$H_s$ respectively. By substituting this decomposition $H=H_c+H_s$ into
Eq. (\ref{holegreen}) and taking momentum conservation into account,
we find that the hole Green function can be written as a 
sort of convolution between a holon ($G_h$) and a spinon ($Z$) term:
\begin{equation}\label{spincharge}
G(p,t)=\int{dQ\over 2\pi} G_h(p-Q,t)\,Z_p(Q,t)
\end{equation}
where $G_h(k,t)$ is just a free propagator:
${\rm Im} \,G_h(k,\omega) =\pi\,\delta(\omega -\epsilon_h(k) )$.
This simple form of the holon Green function is due
to the presence of a gap in the charge excitation spectrum of the model
which confines the low energy processes in the single holon sector.
Instead, the spinon contribution $Z$ is highly non-trivial due to the 
presence of gapless excitations and exhibits anomalous exponents at particular 
momenta $Q_{\nu}$. As a result, the most relevant
singularity in the full spectral function occurs at frequencies determined by
the hole  dispersion: 
\begin{equation}\label{singular}
A(p,\omega)={1\over\pi} {\rm Im} \, G(p,\omega)\propto\left [
\omega-\epsilon_h(p-Q_{\nu})\right ]^{2X_{\nu}(p)-1}
\end{equation}
Here $X_{\nu}(p)$ is a {\it momentum dependent} 
critical exponent determined by the low energy properties of the spinon 
dynamics, while $Q_{\nu}= (2 \nu+1) Q_F$ is an odd multiple of the spinon 
Fermi momentum $Q_F=\pi(1/2-m)$, $m$ being the magnetization  per site. 
Equation (\ref{spincharge}) generalizes the exact form found in the 
$U\to\infty$ limit of the half filled Hubbard model \cite{ssap} where 
$Z_p(Q,t)$ was explicitly calculated and turned out to be independent of 
$p$ and $t$. In this limit, the hole dispersion is 
$\epsilon_h(k)=-2\cos k$ and the critical exponent takes the value 
$X_0=X_{-1}=1/4$. 

In the following, we will analyze the long wavelength behavior of 
$Z_p(Q,t)$ which leads to Eq. (\ref{singular}) in the
particular case of single (spin down) hole in the $tJ$ model at arbitrary
magnetization $m$. The case $m=0$ gives information about the Mott
insulator while the $m>0$ ($m<0$) choice refers to photoemission
(inverse photoemission) experiments  in ``superconductors" via spin-up
(spin-down) particle-hole transformation in the less than half filled 
attractive
Hubbard model. Due to the universality underlying the behavior of one 
dimensional physics,  we expect that these results will be qualitatively valid 
for generic 1D electron system displaying a gap in the excitation spectrum.
In fact, it is well known \cite{solyom} that in this case the renormalization
group flow drives the model towards the Luther Emery fixed point 
irrespective of the details of the microscopic hamiltonian. On the other hand, 
the $tJ$ model allows for a direct comparison with Lanczos diagonalizations 
which can be pushed to fairly large lattice size in such a system.

The single hole problem in the $tJ$ model can be reduced to a pure 
spin problem by a Galileo transformation 
\cite{zhong} which fixes the hole at the origin $O$ of the $L$-site
lattice ($L$ is chosen to be even). 
In this way, the charge degree of freedom can 
be exactly traced out leaving the problem of an effective 
momentum dependent spin  hamiltonian:
\begin{equation}\label{tj}
H^{eff}_p \,=\, 
-\left [ e^{i p } T + e^{-ip} T^{\dagger}\right ] + 
J \sum\limits_{i=1}^{L-2} {\bf S}_i \cdot {\bf S}_{i+1} .
\end{equation}
Here $p$ is the total lattice momentum of the one hole state and 
$T$ is the translation operator along the squeezed chain with $l=L-1$
sites, i.e. without the origin $O$. The hamiltonian $H^{eff}$ is
written as the sum of two physically different contributions:
The magnetic term is the standard Heisenberg model with {\it open} boundary 
conditions because no spin is present at the  origin, while the kinetic 
of the hole manifests itself via the action of the translation operator $T$.

This mapping of the hole problem into a spin hamiltonian is exact for every $J$
and allows to interpret the presence of the hole as the inclusion of a special
type of {\it boundary operator} in the bulk spin hamiltonian.  
An insight to the general features of the energy spectrum of this hamiltonian
can be obtained by examining the $J \to 0$ limit, where all the 
spin configurations on the squeezed chain which are eigenstates of the
translation operator $T |\psi_Q>= e^{i Q} |\psi_Q>$ are degenerate 
provided they correspond to the same (spinon) momentum $Q$.
This degeneracy is lifted by the magnetic term, which, at first 
order in $J$, selects the lowest energy state of the 
Heisenberg ring with the given spinon momentum $Q$. 
The corresponding hole energy is $E=\epsilon_h(p-Q)+ 
\epsilon_s(Q)$, where the holon band is $\epsilon_h(k)=-2\cos k$ 
and the $O(J)$ spinon dispersion $\epsilon_s(Q)$ only depends on the
bulk properties of the Heisenberg model. In this limit, the effects
of spin-charge decoupling on the energy spectrum come out rather naturally 
as well as the role of the hole kinetic contribution in modifying the 
boundary conditions of the Heisenberg model, from open to periodic.
Due to the peculiar form of the hole boundary operator in Eq. (\ref{tj}),
the long wavelength behavior of $H^{eff}$ is associated to a new class 
of fixed points different from those found in the framework of the 
{\it static} impurity problem \cite{affleck}.

After a standard  Jordan-Wigner transformation 
the spin hamiltonian maps onto   an electron system of interacting 
spinless fermions at density $\rho={1\over 2}-m$. 
At low energy the relevant degrees of freedom for a many fermion
system are  the momenta close to the Fermi points $\pm Q_F$.
It is then possible to take the continuum limit of the model defining
two independent fermionic fields $\psi_R (x)$ and $\psi_L(x)$ \cite{review}
for the right $k\sim Q_F$ and left $k \sim -Q_F$ movers on the 
squeezed  chain  with  $0 <x < l$.
The long wavelength limit of the translation operator can be written 
in terms of a well defined  spinon momentum operator $\hat P$: 
$T=e^{i \hat P}$ where 
\begin{eqnarray}
\hat P=Q_F &&\int\limits_0^l dx \,\Big\{
\left [\psi^{\dagger}_R (x) \psi_R(x) -  \psi^{\dagger}_L (x) \psi_L (x)
\right] +\nonumber\\
&i& \left[ \psi_R^{\dagger} (x)  \partial_x \psi_R (x) 
+ \psi_L^{\dagger} (x) \partial_x \psi_L (x) \right]\Big\}
\label{momentum}
\end{eqnarray}
As anticipated, the competition between the hole kinetic term and 
the magnetic interaction in $H^{eff}$ 
selects a particular effective boundary condition for the
interacting fermion  gas which simulates the presence of a scattering 
potential at the boundary together with a magnetic flux across the ring.
This gives rise to independent boundary conditions for the 
right and left movers defined by two phase shifts:
\begin{eqnarray} \label{bcondp}
\psi_R^{\dagger} (x+l) &=&  e^{ i  \delta_R } \psi^{\dagger}_R (x) 
\nonumber  \\ 
\psi_L^{\dagger} (x+l) &=&  e^{ i  \delta_L } \psi^{\dagger}_L (x)  
\end{eqnarray}  
The long wavelength analysis proceeds by noting that, 
due to spin charge decoupling, the 
effective hamiltonian splits into the sum of two
{\it commuting} terms 
\begin{equation}
H^{eff}=\epsilon_h(p -\hat P)+ H_J
\label{split}
\end{equation}
governing charge and spin dynamics respectively. Here $H_J$ is the usual 
long wavelength form of the Heisenberg 
hamiltonian in terms of the fermionic fields ($\psi_L$, $\psi_R$)
\cite{review} with the particular boundary conditions (\ref{bcondp}).
It is known that the interacting hamiltonian (\ref{split}) can be turned into
a free Fermi problem by a canonical transformation \cite{review}
which further modifies the boundary conditions (\ref{bcondp}). 
The bulk Luttinger liquid properties of $H_J$ are described by
the dressed charge $K_\rho$  shown in Fig. 1 \cite{korepin}.

Having characterized the long wavelength physics of the effective hole
hamiltonian (\ref{tj}) in terms of the dressed charge and the two phase shifts, 
let us move to the analysis of the spectral properties of the hole motion. 
As a first step, we fix our attention on the {\it overlap} 
$\zeta_p$ between the pure Heisenberg ground state on $L$ sites 
$|\Psi_0>$ and the one hole ground state $|\Psi_p>$ at fixed momentum $p$: 
\begin{equation}\label{zeta}
\zeta_p= |<\Psi_p|c_{p,\downarrow}|\Psi_0>|^2\,\propto \,L^{-2X_{\nu}(p)}
\end{equation}
This quantity naturally enters the calculation of the hole spectral function 
as can be immediately checked by use of Lehmann representation.
As indicated in Eq. (\ref{zeta}), the overlap $\zeta_p$ vanishes
in the thermodynamic limit with a critical exponent which can be
explicitly evaluated in terms of the previously introduced phase shifts:
\begin{equation}
X_{\nu}(p)= 
K_{\rho} \left({\delta_R+\delta_L\over 2\pi}+\nu\right )^2 + 
{1\over 4 K_{\rho}} \left({\delta_R-\delta_L\over 2\pi}\right )^2 
\label{exponent}
\end{equation}
The calculation parallels the known derivation of the orthogonality 
catastrophe in the impurity problem \cite{pwa}.
Here, the integer number $\nu$ defines the total spinon momentum of the
intermediate states $ Q_{\nu} = (2 \nu+1) Q_F$  
corresponding to an odd number of low energy spinons. 

The phase shifts (\ref{bcondp}) are non universal  depending on the short
wavelength properties of the model, however it is possible to relate them to 
the form of the energy spectrum $E(p)$ of one hole at fixed momentum $p$
which, according to Eq. (\ref{split}), can be written as a sum 
of a charge and a spin contribution. In fact,
we expect that low energy spinon dynamics is governed by some effective 
conformal field theory which should reflect on the structure of
the size corrections to the {\it spinon contribution} to the ground
state energy: $E(p) - L\,\epsilon_{\infty}(p) = \left ( \Delta E_c(p) + 
\Delta E_s (p)\right )/L$ with
\begin{equation}
\Delta E_s (p) = -v_s\,{\pi\over 6} \,+ \,2\pi v_s\,X_{\nu}(p)
\label{conformal} 
\end{equation}
Here $v_s$ is the spinon velocity: $v_s=d\epsilon_s(Q)/dQ$ evaluated at $Q_F$
and $X_{\nu}(p)$ coincides with the critical exponent (\ref{exponent}).

In order to determine the unknown phase shifts $\delta_R$ and $\delta_L$,
we have analyzed the size corrections to the ground state
energy (at fixed momentum $p$) in the Bethe ansatz soluble models: The Hubbard
model  at $U > 0$ \cite{liebwu} and the $tJ$ model at $J=2$ \cite{bares} with 
one hole and arbitrary magnetization. By suitably generalizing 
the pioneering work of Woynarovich \cite{woynarovich} to the 
single hole case, 
we found exactly the form (\ref{conformal}) of the 
energy size corrections, with quantitative predictions for the phase shifts
which in fact explicitly depend on the
total momentum $p$ at every non-zero magnetization $m$. For 
$m=0$, i.e. for the Mott insulator case, instead, we always find that 
only one of the two phase shifts is different from zero and takes the value 
$\pi$, both in the Hubbard and in the $tJ$ model.
Another analytic limit is the $J\to 0$ at arbitrary magnetization $m$ where
again the phase shifts are independent of $p$ but are functions of the 
magnetization: $\delta_R=\pi\,(1-m)$ and $\delta_L=\pi\, m$. 
Figure 2 shows the exponent $X_{0}(p)$ as a function of the total 
momentum $p$ for the Bethe ansatz solvable
limit of the $tJ$ model ($J=2$) at several magnetizations. A comparison
with the value of the overlap exponent obtained by use of Eq. (\ref{zeta})
through Lanczos diagonalization of the model with magnetization $m=\pm 0.25$
is shown in Fig. 3 . 
Lattice sizes ranging from 16 up to 32 sites have been used to fit the
exponent $2X_{\nu}(p)$ in Eq. (\ref{zeta}) leading to a quite good agreement 
between analytical and numerical results. This comparison gives confidence 
on the interpretation of the Bethe ansatz results for the single hole size 
correction in the framework of conformal field theory. 

Now we are ready to use the previous analysis 
for the evaluation of the hole spectral function.
In fact, the spinon contribution to the Green function $Z_p(Q,t)$ 
appearing in (\ref{holegreen}) can be calculated within the 
described  formalism leading to the expression:
\begin{equation}
Z_p(R,t)=<\Psi_0| e^{ i (\hat P R - H_J t +E_0 t) } |\Psi_0>
\label{formal}
\end{equation}
The dependence on the total momentum $p$ occurs only through the phase shifts 
(\ref{bcondp}) defining the boundary condition to $H_J$. The asymptotic
behavior of $Z_p(R,t)$ can be analytically evaluated as:
\begin{equation}
Z_p(R,t) \sim {e^{i Q_{\nu}
R}\over (R-v_st)^{X_{\nu}(p)+\Delta}(R+v_st)^{X_{\nu}(p)-\Delta}} 
\label{factor}
\end{equation}
showing that singularities characterized by different exponents 
(\ref{exponent}) occur at wavevectors $Q_{\nu}=(2\nu+1)Q_F$.
The additional critical exponent $\Delta$ can be expressed in terms of 
the phase shifts:
$\Delta=(\delta_R+\delta_L+2\pi\nu) (\delta_R-\delta_L)/(2\pi)^2$.
When the Fourier transform of Eq. (\ref{factor}) is substituted into
the asymptotic form of the Green function (\ref{spincharge}) we
obtain the anticipated expression (\ref{singular}) which
constitutes the main result of this Letter together with the 
analytical evaluation of the critical exponent $x_{\nu}(p)$ in Bethe
ansatz soluble models. 
Due to the allowed values for $\nu$ in (\ref{exponent}) singularities
at  $Q_{\nu}=(2 \nu+1) Q_F$ are predicted, determining ``shadow bands'' 
in the spectral function. 

From a physical point of view, our results
show that the spectral function of Mott insulators and superconductors
is characterized by branch cut singularities with exponents depending,
in the latter case, on the momentum $p$ of the injected particle.
This feature is shared by all the models we have investigated:
The Bethe ansatz solvable Hubbard and $tJ$ models and the $J\to 0$ limit
of the $tJ_{XY}$ model \cite{extended}. We believe that it is
a general feature of hole motion in 1D correlated systems thereby
providing definite predictions for the analysis of
photoemission experiments in quasi one dimensional systems 
characterized by a gap either in the charge or in the spin spectrum.

It is a pleasure to thank M. Fabrizio for extensive discussions and to
acknowledge kind hospitality at SISSA (AP), Cantoblanco University (SS) 
 and ISI foundation (EU Contract ERBCHRX-CT920020).

\begin{figure}
\caption{Correlation exponent (or dressed charge) $K_{\rho}$
for the Heisenberg model as a function of magnetization. The line is
obtained by numerical solution of the integral equation for
$\xi_{22}$ of Ref. \protect\cite{korepin}.}
\label{fig1}
\end{figure}

\begin{figure}
\caption{Lowest critical exponent $X_{\nu}(p)$ as a function of the total 
momentum $p$ of 
the spin down hole in the $tJ$ model at $J=2$ with several magnetizations $m$.
Results are obtained by numerical solution of the set of integral equations
 defining  the correction to scaling of the one hole ground state
energy \protect\cite{extended}}
\label{fig2}
\end{figure}

\begin{figure}
\caption{Comparison between the analytical results of Fig. 2 for $m=\pm 0.25$
and Lanczos diagonalization. In the latter case, the exponent $X_{\nu} (p)$ 
is obtained by a size scaling of the numerical evaluation of the overlap
$\zeta_p$ as defined by Eq. (\protect\ref{zeta}). 
Solid line: analytical results,
open dots: Lanczos data for $m=0.25$, full dots: Lanczos data for $m=-0.25$.
}
\label{fig3}
\end{figure}

\end{document}